\documentclass[11pt]{article}
\usepackage[utf8]{inputenc}
\usepackage{amsfonts}
\usepackage{amsmath}
\usepackage{amssymb}
\usepackage{indentfirst}
\usepackage{graphicx}
\usepackage[colorlinks]{hyperref}
\usepackage{cite}
\usepackage{xcolor}
\usepackage{soul} 
\usepackage{microtype}
\usepackage[normalem]{ulem}

\numberwithin{equation}{section}
\allowdisplaybreaks

\begin{document}

\begin{titlepage}
\vspace{3cm}

\baselineskip=24pt

\begin{center}
\textbf{\LARGE{Hypersymmetric extensions of Maxwell Chern-Simons gravity in (2+1) dimensions}}
\par\end{center}{\LARGE \par}

\begin{center}
	\vspace{1cm}
	\textbf{Ricardo Caroca}$^{\ast}$,
	\textbf{Patrick Concha}$^{\ast}$,
    \textbf{Javier Matulich}$^{\ddag}$,
	\textbf{Evelyn Rodríguez}$^{\dag}$,\\
	\textbf{David Tempo}$^{\star}$
	\small
	\\[5mm]
    $^{\ast}$\textit{Departamento de Matemática y Física Aplicadas, }\\
	\textit{ Universidad Católica de la Santísima Concepción, }\\
\textit{ Alonso de Ribera 2850, Concepción, Chile.}
	\\[2mm]
	$^{\ddag}$\textit{Université Libre de Bruxelles and International Solvay Institutes,}\\
	\textit{ULB-Campus Plaine CP231, B-1050 Brussels, Belgium, }
	\\[2mm]
	$^{\dag}$\textit{Departamento de Física, Universidad del Bío-Bío, }\\
	\textit{Avenida Collao 1202, Casilla 5-C, Concepción, Chile} \\[2mm]
	$^{\star}$\textit{Departamento de Ciencias Matemáticas y Físicas, Facultad de Ingeniería}\\
	\textit{Universidad Católica de Temuco, Chile.}\\[5mm]
	\footnotesize
	\texttt{rcaroca@ucsc.cl},
	\texttt{patrick.concha@ucsc.cl},
    \texttt{javier.matulich@ulb.be},
	\texttt{ekrodriguez@ubiobio.cl},
	\texttt{jtempo@uct.cl}
	\par\end{center}
\vskip 26pt
\begin{abstract}
We present a consistent way of coupling three-dimensional Maxwell Chern-Simons gravity theory with massless spin-$\frac{5}{2}$ gauge fields. We first introduce the simplest hyper-Maxwell Chern-Simons gravity containing two massless spin-2 fields coupled with a massless Majorana fermion of spin-$\frac{5}{2}$ and whose novel underlying 
superalgebra is explicitly constructed. Then, we present three alternative hypersymmetric extensions of the Maxwell algebra  which are shown to emerge from the Inönü-Wigner contraction procedure of precise combinations of the $\mathfrak{osp}\left(1|4\right)$ and the $\mathfrak{sp}\left(4\right)$ algebras. This allow us to construct distinct types of hyper-Maxwell Chern-Simons theories that extend to include generically interacting non-propagating spin-$4$ fields accompanied by one or two spin-$\frac{5}{2}$ gauge fields.
\end{abstract}
\end{titlepage}\newpage {} {\baselineskip=12pt \tableofcontents{}}

\section{Introduction}

Hypergravity is an early alternative to supergravity theory \cite{Freedman:1976xh,Deser:1976eh} proposed by Aragone and Deser \cite{Aragone:1979hx},
that deems a spin-$5/2$ field as the superpartner of the graviton. Although this
proposal initially attracted some attention \cite{Aragone:1979hx,Berends:1979kg,Berends:1979wu,Aragone:1980rk},
it was promptly discarded due to the incompatibility of the minimal coupling between gravity (spin-$2$) and the spin-$5/2$ field with higher spin (HS) gauge invariance.  This obstruction relies on the fact that the HS gauge variation
of the Einstein-Hilbert action is proportional to the Ricci tensor
so that it cannot be cancelled by means of the minimally coupled spin-$5/2$
field, which is instead proportional to the full Riemann tensor. Nonetheless,
due to the particular  relationship between the Riemann tensor and the
Ricci tensor in three spacetime dimensions, Aragone and Deser managed
to formulate the first consistent interacting HS theory \cite{Aragone:1983sz}. The sought after extension was later
constructed for any dimensions by Fradkin and Vasiliev with some crucial
requirements as the inclusion of a negative cosmological constant,
and an infinite tower of HS fields, in order to successfully evade
the no-go theorems \cite{Fradkin:1987ks,Fradkin:1986qy,Vasilev:2011xf,Vasiliev:1990en}
(for recent reviews see e.g., \cite{Vasiliev:1995dn,Bekaert:2005vh,Bekaert:2010hw,Sagnotti:2013bha}).
It is noteworthy that the presence of a negative cosmological constant
in these HS gauge theories triggered sharp increase of interest in
the subject, due to its salient role in the AdS/CFT correspondence
\cite{Maldacena:1997re,Gaberdiel:2010pz,Creutzig:2011fe,Candu:2012jq},
as well as its close relationship with the tensionless limit of string
theory (see e.g., \cite{Sundborg:2000wp,Sezgin:2002rt,Taronna:2010qq,Sagnotti:2010at,Gaberdiel:2014cha,Rahman:2015pzl}).

Higher spin gauge theories in three spacetime dimensions have shown
to be very useful laboratories to investigate many intricate features of their four and higher dimensional counterparts. In $2+1$ dimensions,
anti-de Sitter hypergravity theory was studied in \cite{Chen:2013oxa,Zinoviev:2014sza,Henneaux:2015ywa}.
The theory is constructed in terms of a Chern-Simons (CS) action for
two copies of the $\mathfrak{osp}(4|1)$ superalgebra \footnote{Along these lines,  (super)conformal gravity formulated as a CS gauge theory for  ($\mathfrak{osp}\left(1|4\right)$) $\mathfrak{sp}\left(4\right)$ (super)algebra was done in \cite{VanNieuwenhuizen:1985ff,Rocek:1985bk,Horne:1988jf,Lindstrom:1989eg,Nishino:1991sr}, while the corresponding asymptotic structure has been studied in \cite{Afshar:2011qw,Bertin:2012qw,Afshar:2013bla,Fuentealba:2020zkf}.}, and contains
a spin-$2$ field, a spin-$4$ field and a spin-$5/2$ field. The
asymptotic structure analysis of this model was performed in \cite{Henneaux:2015ywa}
where it was shown that its asymptotic symmetry algebra is given by
two copies of the hypersymmetric extension of the $W(2,4)$ algebra,
known as $WB_{2}$-superalgebra \cite{Bellucci:1994xa}, and $W\left(2,5/2,4\right)$-superalgebra
\cite{Figueroa-OFarrill:1991huu}. Furthermore, this study also revealed the existence
of hypersymmetry bounds involving a nonlinear function of the mass,
angular momentum and bosonic higher spin charges, as well as an interesting
class of HS solutions including solitons and extremal HS black holes with unbroken hypersymmetries \cite{Henneaux:2015ywa,Henneaux:2015tar}.In the vanishing cosmological constant
limit it was verified that the spin-$4$ field decouples thus reproducing
the hypergravity theory studied in \cite{Fuentealba:2015jma,Fuentealba:2015wza}.

Recently, a novel extension of the Poincaré algebra with half-integer
spin generators in any dimension allowed to reformulate the hypergravity
theory of Aragone and Deser as a genuine gauge theory in terms of
a CS action\footnote{Fermionic HS generalizations of Poincar\'e superalgebra were previously studied in \cite{Hietarinta:1975fu} for any dimension.} \cite{Fuentealba:2015jma,Fuentealba:2015wza}. The asymptotic structure analysis led to obtain a nonlinear hypersymmetric
extension of the $\mathfrak{bms}_{3}$ algebra, along with nonlinear
bounds for the energy. Moreover, despite this theory involves HS fields, the absence of bosonic HS fields allows to describe
the theory in standard Riemann-Cartan geometry. On the other hand, the fact that CS forms possess natural generalizations
in odd dimensions (see e.g. \cite{Banados:1996hi}) suggests that this
construction can be carried out by means of the higher dimensional
hyper-Poincaré algebras in \cite{Fuentealba:2015jma}. Indeed, this
was done in \cite{Fuentealba:2019bgb}, where it was shown that a
spin-5/2 field can be consistently coupled to gravity without cosmological
constant in five dimensions, where the gravitational sector is described
by the quadratic Gauss-Bonnet action, and thereby the field equations
are of second order for the metric. It is worth emphasizing that the
gauge symmetries close for a finite extension of the hyper-Poincaré
algebra that admits an invariant trilinear form thus allowing to
formulate the theory by means of a five-dimensional CS form.

A natural question that arises is whether it is possible to construct consistent extensions of the hypergravity theories based on extensions of hyper-Poincaré
algebra. It should be mentioned that in order to deepen into this possibility,
it is mandatory to have full control on the building blocks, namely,
a consistent algebra with a non-degenerate invariant tensor. In what
follows we address this question by considering a nontrivial extension
of the Poincaré algebra known long ago as the Maxwell algebra \cite{Bacry:1970ye,Bacry:1970du},
being associated with the symmetry group of Dirac (Klein-Gordon) equation
minimally coupled to a constant electromagnetic field in Minkowski
space in $3+1$ dimensions \cite{Schrader:1972zd}. In any dimension, this algebra is characterized by the commutator
\begin{equation*}
    \left[ P_{a},P_{b}\right] =Z_{ab}\,,
\end{equation*}
modifying in this way the commutator of the momentum generators, which vanishes for the Poincar\'e algebra. The study of different aspects concerning the Maxwell algebra in four and
higher dimensions including its derivation as an S-expansion on the
AdS algebra can be found in \cite{Salgado:2014qqa,Izaurieta:2006zz}.  The Maxwell group symmetries and its generalizations have been useful to extend standard General Relativity through CS and Born-Infeld gravity theories in odd and even spacetime dimensions, respectively \cite{Edelstein:2006se,Izaurieta:2009hz,Concha:2013uhq,Concha:2014vka}. Deformations of this algebra and their dynamics through non-linear realizations have been investigated \cite{Bonanos:2008ez,Bonanos:2008kr,Gomis:2009dm}, as well as other interesting applications, see e.g., \cite{deAzcarraga:2010sw,Durka:2011nf,deAzcarraga:2012qj,Salgado-Rebolledo:2019kft,Kibaroglu:2020tbr,Cebecioglu:2021dqb}.

In three spacetime dimensions, Maxwell CS gravity appears as a very appealing alternative theory of gravity in vacuum introduced in\footnote{A CS gravity theory based on Maxwell algebra in $2+1$ was initially considered in \cite{Cangemi:1992ri, Duval:2008tr} as a prominent model leading to the two-dimensional linear gravities referred in    \cite{Fukuyama:1985gg,Isler:1989hq,Chamseddine:1989yz,Cangemi:1992bj,Cangemi:1993sd} by a dimensional reduction.} \cite{Cangemi:1992ri, Duval:2008tr, Salgado:2014jka} and subsequently studied in \cite{Hoseinzadeh:2014bla,Concha:2018zeb,Adami:2020xkm}.  The asymptotic structure of the Maxwell CS theory was investigated in \cite{Concha:2018zeb} by imposing a set of suitable boundary conditions resulting in an asymptotic symmetry algebra given by a deformation of the  $\mathfrak{bms}_{3}$ algebra, known to emerge from the asymptotic symmetry analysis of General Relativity at null infinity \cite{Ashtekar:1996cd, Barnich:2006av, Barnich:2010eb,Barnich:2014cwa,Fuentealba:2017omf}, which goes in line with the result obtained in \cite{Caroca:2017onr} by expanding the Virasoro algebra (see also \cite{Parsa:2018kys, Concha:2019eip}).  Interestingly, the presence of the gravitational Maxwell gauge field modifies not only the asymptotic symmetry but also the vacuum of the theory \cite{Concha:2018zeb}. Physical implications of the gravitational Maxwell gauge field have also been explored in the context of spin-3 gravity \cite{Caroca:2017izc}, non-relativistic gravity \cite{Aviles:2018jzw,Concha:2019mxx,Concha:2020sjt,Concha:2020ebl} and supergravity \cite{Concha:2018jxx,Concha:2019icz}. It is worth noting that in three-dimensions the extension of Poincaré algebra found by Hietarinta in \cite{Hietarinta:1975fu} becomes isomorphic to three-dimensional Maxwell algebra that amounts to a simple interchanging of roles between the translation generators $P_a$ and  $Z_a$  in Maxwell algebra \cite{Bansal:2018qyz,Chernyavsky:2019hyp,concha:2021aaa} . Indeed, both CS gravity theories based on Hietarinta and Maxwell algebras have been explored in \cite{Bansal:2018qyz,Chernyavsky:2020fqs} showing in particular that upon spontaneous breaking of a local symmetry lead precisely to the topologically massive gravity theory \cite{Deser:1981wh} and the minimal massive gravity\cite{Bergshoeff:2014pca}.

One of the main advantages of working in three spacetime dimensions is that the Maxwell algebra can alternatively be recovered as an Inönü-Wigner (IW) contraction of three copies of the $\mathfrak{so}\left(2,1\right)$ algebra which also provides a nontrivial invariant form that is imperative for the construction of extensions of Maxwell CS gauge theories \cite{Concha:2018jjj}. In particular, the spin-3 Maxwell algebra as well as its  corresponding invariant bilinear form can be derived by contracting three copies of the $\mathfrak{sl}\left(3,\mathbb{R}\right)$ symmetry \cite{Caroca:2017izc} \footnote{In  \cite{Campoleoni:2010zq,Gonzalez:2013oaa,Matulich:2014hea,Gary:2014ppa} following suitable IW contractions have allowed to formulate the three-dimensional higher spin gravity with vanishing cosmological constant as a CS gauge theory likewise, as their well-known counterparts on $AdS_{3}$ \cite{Henneaux:2010xg,Campoleoni:2010zq,Campoleoni:2011hg,Henneaux:2013dra,Bunster:2014mua}.}. One may then ask whether a hypersymmetric extension of the Maxwell algebra with fermionic spin-$\frac{5}{2}$ generators, which transform in an spin-$\frac{3}{2}$ irreducible representation of the Lorentz group, can be obtained by contracting diverse combinations of the $\mathfrak{osp}\left(1|4\right)$ and the $\mathfrak{sp}\left(4\right)$ algebras. In this work, we show that not one but three distinct hypersymmetric extensions of the Maxwell algebra, including their invariant bilinear forms, can be effectively derived through the IW procedure. The obtained hyper-Maxwell algebras indeed require the presence of spin-4 generators and allow us to construct CS hypersymmetric gravity theories. Furthermore, we show also that a remarkable hyper-Maxwell CS gravity theory without spin-4 gauge fields can be constructed, and whose underlying superalgebra results to be a subsuperalgebra of one of the hyper-Maxwell algebras that include spin-4 generators, which transform in an spin-3 irreducible representation of the Lorentz group.

The paper is organized as follows: In Section 2, we briefly review the Maxwell CS gravity theory defined on three spacetime dimensions. Sections 3, 4 and 5 contain
our main results. In Section 3, we present the simplest CS hypergravity theory invariant under a hypersymmetric extension of the Maxwell algebra. In Section 4, we introduce three alternative hyper-Maxwell algebras including spin-4 generators by considering the IW contraction procedure. Section 5 is devoted to the construction of the CS hypersymmetric gravity theories based on the aforementioned hyper-Maxwell symmetries. Section 6 concludes our work with some discussions about future developments.


\section{Three-dimensional Maxwell Chern-Simons gravity theory}
In this section, we briefly review the three-dimensional Maxwell
CS gravity theory \cite{Cangemi:1992ri, Duval:2008tr, Salgado:2014jka} (see also \cite{Hoseinzadeh:2014bla,Concha:2018zeb,Concha:2018jxx}). This alternative theory of gravity is based on the so-called Maxwell algebra, which can be seen as an extension and deformation of the
Poincaré algebra $\mathfrak{iso}\left(2,1\right)$, which turns out to be a non semi-simple group. In addition to the usual local rotations $J_a$ and local translation generators $P_a$, the Maxwell symmetry is characterized by the presence of three additional Abelian generators $Z_a$. In particular, the Maxwell generators satisfy the following non-vanishing
commutation relations 
\begin{eqnarray}
\left[ J_{a},J_{b}\right] &=&\epsilon _{~ab}^{c}J_{c}\,,  \qquad \,
\left[ J_{a},P_{b}\right] =\epsilon _{~ab}^{c}P_{c}\,,  \notag \\
\left[ J_{a},Z_{b}\right] &=&\epsilon _{~ab}^{c}Z_{c}\,,  \qquad
\left[ P_{a},P_{b}\right] =\epsilon _{~ab}^{c}Z_{c}\,,  \label{Max}
\end{eqnarray}%
where $a,b,c=0,1,2$ are Lorentz indices which are lowered and raised with
the Minkowski metric $\eta _{ab}=\left( -1,1,1\right) $ and $\epsilon _{abc}$
is the three-dimensional Levi Civita tensor which satisfies $\epsilon
_{012}=-\epsilon ^{012}=1$.

The most general quadratic Casimir invariant for the Maxwell algebra is \cite{Cangemi:1992ri, Duval:2008tr, Salgado:2014jka}
\begin{equation}
    C=\alpha_{0}J^aJ_a+\alpha_{1}P^aJ_a+\alpha_{2}\left(P^aP_a+J^aZ_a\right)\, , \label{CasMax}
\end{equation}
where $\alpha _{0}$, $\alpha _{1}$, and $\alpha _{2}$ are arbitrary constants. Then, the Maxwell algebra admits the following non-vanishing components of
the invariant tensor,
\begin{eqnarray}
\left\langle J_{a}J_{b}\right\rangle &=&\alpha _{0}\eta _{ab}\,,  \qquad
\left\langle J_{a}P_{b}\right\rangle =\alpha _{1}\eta _{ab}\,,  \notag \\
\left\langle J_{a}Z_{b}\right\rangle &=&\alpha _{2}\eta _{ab}\,,  \qquad
\left\langle P_{a}P_{b}\right\rangle =\alpha _{2}\eta _{ab}\,.  \label{ITM}
\end{eqnarray}

The gauge connection one-form $A$ can be conveniently chosen as follows \footnote{More general choices of the gauge field can be considered leading to a more general Maxwell-CS theory that includes a non vanishing torsion term.}
\begin{equation}
A=\omega^{a}J_{a}+e^{a}P_{a}+k^{a}Z_{a}\,,  \label{1f}
\end{equation}
where $\omega^{a}$ is the (dualized) spin-connection, $e^{a}$ denotes the dreibein and $k^{a} $ is the so-called gravitational Maxwell gauge field. The
corresponding curvature two-form is given by
\begin{equation}
F=R^{a} J_{a}+T^{a} P_{a}+F^{a}
Z_{a}\,,
\end{equation}
where
\begin{eqnarray}
R^{a} &=& d\omega^{a}+\frac{1}{2}\epsilon _{~bc}^{a}\omega^{b}\omega^{c}\,,
\notag \\
T^{a} &=& de^{a}+\epsilon _{~bc}^{a}\omega^{b}e^{c}\,,  \notag
\\
F^{a} &=& dk^{a}+\epsilon _{~bc}^{a}\omega^{b}k^{c}+\frac{1}{2}
\epsilon _{~bc}^{a}e^{b}e^{c}\,.  \label{MaxCurv}
\end{eqnarray}

Then, considering the gauge-connection (\ref{1f}) and the invariant tensor (\ref{ITM}), the corresponding action for the Maxwell algebra can be described in terms of the three-dimensional CS action,
\begin{equation}
I_{CS}=\frac{k}{4\pi }\int \left\langle AdA+\frac{2}{3}A^{3}\right\rangle \,,
\label{CS}
\end{equation}
with $k=\frac{1}{4G}$ being the CS level of the theory related to the gravitational
constant $G$. Indeed, the action reads \cite{Cangemi:1992ri, Duval:2008tr, Salgado:2014jka}  (see also \cite{Hoseinzadeh:2014bla,Concha:2018zeb,Concha:2018jxx}):
\begin{equation}
I_{\text{Maxwell}} =\frac{k}{4\pi }\int 2\alpha
_{1}R^{a}e_{a} +\alpha _{2}\left( e^{a}T_{a}+2R^{a}k_{a} \right) +\alpha _{0} L(\omega) \,,  \label{MaxCS}
\end{equation}

where

\begin{equation}
    L(\omega)= \left( d\omega^{a}+ \frac{1}{3}\epsilon _{~bc}^{a}\omega^{b}\omega^{c}\right)\omega_{a} \, , \label{LorCS}
\end{equation}
is the Lorentz-Chern-Simons form.

The Maxwell CS action contains three independent sectors proportional to $
\alpha _{0}$, $\alpha _{1}$, and $\alpha _{2}$. The parity-odd term given by the Lorentz CS three form \cite{Achucarro:1987vz,Witten:1988hc} appears along the $\alpha _{0}$ constant while the Einstein-Hilbert term is related to the $\alpha _{1}$
constant. On the other hand, the additional gauge field $k^{a}$ only contributes to the $\alpha_2$ sector. 

In particular, the equations of motion are given by 
\begin{eqnarray}
 \delta \omega_a&:& \qquad \alpha_0 R^a + \alpha_1 T^a + \alpha_2  F^a =0\,, \notag \\
 \delta k_a &:& \qquad \alpha_2 R^a =0\,, \notag \\
 \delta e_a &:& \qquad \alpha_1 R^a + \alpha_1 T^a =0\,. \label{eommaxwell}
\end{eqnarray}
It follows that when $\alpha_2\neq0$, the previous equations can be equivalently written as the vanishing of the curvature two-forms \eqref{MaxCurv}. It is worth mentioning that the suitable choice of the gauge field in (\ref{1f}) allows to describes generically the theory in Riemannian geometry (torsionless). The standard $(2+1)$-dimensional gravity in vacuum is then recovered for case $\alpha_{0}=\alpha_{2}=0$, being the well-known CS theory for $\mathfrak{iso}\left(2,1\right)$ \cite{Achucarro:1987vz,Witten:1988hc}.

In this work, we will “hypersymmetrize” the three-dimensional Maxwell gravity. As we will see, some hypersymmetric extensions will also  require the presence of spin-4 gauge fields. The construction of the simplest hypersymmetric extension of the Maxwell CS gravity is discussed in the next section.

\section{Hyper-Maxwell Chern-Simons gravity theory}

Here we present the simplest hypersymmetric extension of the Maxwell CS gravity theory. To this end, we construct a hyper-Maxwell algebra by introducing fermionic generators which transform in an spin-$\frac{3}{2}$ irreducible representation of the Lorentz group. Therefore, the hyper-Maxwell algebra is spanned by the set $\{J_a,P_a,Z_a,Q_{\alpha a}\}$ whose generators satisfy the following non-vanishing (anti-)commutation relations:
\begin{eqnarray}
\left[ J_{a},J_{b}\right]  &=&\epsilon _{~ab}^{m}J_{m}\,,\qquad \left[
J_{a},P_{b}\right] =\epsilon _{~ab}^{m}P_{m}\,, ,  \notag \\
\left[ J_{a},Z_{b}\right]  &=&\epsilon _{~ab}^{m}Z_{m}\,,\qquad \left[
P_{a},P_{b}\right] =\epsilon _{~ab}^{m}Z_{m}\,,   \notag \\
\left[ J_{a},Q_{\alpha b}\right] &=&\frac{1}{2}\left( \Gamma _{a}\right) _{%
\text{ }\alpha }^{\beta }Q_{\beta b}+\epsilon _{abc}Q_{\beta }^{~c}\,,
 \notag\\
\left\{ Q_{\alpha a},Q_{\beta b}\right\}  &=&-\frac{4}{3}\eta
_{ab}Z_{c} \left( C\Gamma ^{c}\right) _{\alpha \beta }+\frac{5}{3}%
\epsilon _{abc}C_{\alpha \beta }Z^{c}+\frac{2}{3}Z_{\left( a\right. |}\left(
C\Gamma _{|\left. b\right) }\right) _{\alpha \beta }\,,
 \label{hmalgebra}
\end{eqnarray}
where $Q_{\alpha a}$ are $\Gamma$-traceless vector-spinor generators that fulfill $\left(\Gamma^{a}\right)_{\ \alpha}^{\beta}=\Gamma^{a}Q_{a}=0$. Here $C$ is the charge conjugation matrix\begin{equation}
C_{\alpha\beta}=\begin{pmatrix}
0 & -1\\
1 & 0\\
\end{pmatrix}\, ,
\end{equation}
which satisfies $C^{T}=-C$ and $C\Gamma^{a}=(C\Gamma^{a})^{T}$ with $\Gamma^{a}$ being the Dirac matrices in three spacetime dimensions. We shall denote this algebra as $\widehat{\mathfrak{hm}}$ which, similarly to the hyper-Poincaré algebra \cite{Fuentealba:2015jma}, does not require an enlargement of the Lorentz group neither the introduction of bosonic HS generators to satisfy the Jacobi Identities. One can notice that the algebra $\widehat{\mathfrak{hm}}$ has a similar structure to the so-called non-standard Maxwell superalgebra \cite{Lukierski:2010dy,Concha:2020atg} in which the translational generators $P_a$ are not expressed as bilinear expressions of fermionic generators. Nonetheless,  since the subsuperalgebra spanned by the generators $J_{a}$, $Z_{a}$ and $Q_{\alpha a}$  is indeed the hyper-Poincaré algebra, as mentioned before, a simple interchanging of roles between the translation generators $P_{a}$ and $Z_{a}$ relates the superalgebra in \eqref{hmalgebra} with its Hietarinta form in \cite{Hietarinta:1975fu}. Therefore, one naturally expects that gauge theories based on the hyper-Maxwell or Hietarinta version of the algebra will be certainly endowed with  quite  different physical implications, see e.g.  \cite{Bansal:2018qyz,Chernyavsky:2019hyp,concha:2021aaa}. In the present work, we will not consider this last reinterpretation of the theory since our purpose is to study the coupling of three-dimensional Maxwell gravity with massless spin-$\frac{5}{2}$ fields.

Let us consider now the hypersymmetric Maxwell CS action being invariant under the  hyper-Maxwell algebra $\widehat{\mathfrak{hm}}$ \eqref{hmalgebra}.  The CS action can be constructed from the gauge field 
\begin{equation}
    A=e^{a}P_{a}+\omega^{a}J_{a}+k^{a}Z_{a}+\Bar{\psi}^{a}Q_{a}\,,\label{oneformhm}
\end{equation}
whose components are the dreibein, the (dualized) spin connection, the gravitational Maxwell field and a Majorana spin-$\frac{5}{2}$ field. In particular, the Majorana conjugate reads $\bar{\psi}_{a\alpha}=\psi_{a}^{\beta}C_{\beta\alpha}$. The corresponding curvature two-form $F=dA+\frac{1}{2}\left[A,A\right]$ is given by 
\begin{equation}
    F_{\widehat{\mathfrak{hm}}}=T^{a}P_{a}+R^{a}J_{a}+\tilde{F}^{a}Z_{a}+D\Bar{\psi}^{a}Q_{a}\,,
\end{equation}
with
\begin{eqnarray}
\tilde{F}^{a}&=&F^a-\frac{3}{2}i\Bar{\psi}_{b}\Gamma^{a}\psi^{b}\,, \notag \\
D\psi^{a}&=&d\psi^{a}+\frac{3}{2}\omega^{b}\Gamma_{b}\psi^{a}-\omega_{b}\Gamma^{a}\psi^{b}\,, \label{curvahm}
\end{eqnarray}
where $T^a$, $R^a$ and $F^a$ are defined in \eqref{MaxCurv}. The fermionic fields are assumed to be $\Gamma$-traceless, i.e, $\Gamma^{a}\psi_{a}=0$. On the other hand, one can easily check that in addition to $C_0=J^{a}J_a$ and $C_1=P^{a}J_a$, the hyper-Maxwell algebra $\widehat{\mathfrak{hm}}$ admits another quadratic Casimir invariant given by
\begin{eqnarray}
C_2&=&P^{a}P_a+J^{a}Z_{a}+Q^{a}_{\alpha}C^{\alpha\beta}Q_{\beta a}\,.
\end{eqnarray}
Hence, the $\widehat{\mathfrak{hm}}$ algebra admits the following non-vanishing components of an invariant tensor,
\begin{eqnarray}
\left\langle J_{a}J_{b}\right\rangle  &=&\alpha _{0}\eta _{ab}\,,\qquad
\left\langle J_{a}P_{b}\right\rangle =\alpha _{1}\eta _{ab}\,,  \notag \\
\left\langle J_{a}Z_{b}\right\rangle  &=&\alpha _{2}\eta _{ab}\,, \qquad
\left\langle P_{a}P_{b}\right\rangle =\alpha _{2}\eta _{ab}\,,  \notag \\
\left\langle Q_{\alpha a}Q_{\beta b}\right\rangle  &=&2\alpha _{2}\left(
\frac{2}{3}C_{\alpha \beta }\eta _{ab}-\frac{1}{3}\epsilon _{abc}\left(
C\Gamma ^{c}\right) _{\alpha \beta }\right) \,,
\label{invtenhm} 
\end{eqnarray}
where $\alpha _{0}$, $\alpha _{1}$ and $\alpha _{2}$ are arbitrary constants. The invariance of such bilinear form under the action of the hyper-Maxwell algebra requires that $\langle J_aZ_b\rangle$, $\langle P_aP_b\rangle$ and $\langle Q_{\alpha a}Q_{\beta b}\rangle$ have the same global coefficient. Then, using the gauge connection one-form \eqref{oneformhm} and the invariant bilinear form \eqref{invtenhm} in the general form of a three-dimensional CS action \eqref{CS}, it reduces, up to boundary terms, to
\begin{eqnarray}
I_{\widehat{\mathfrak{hm}}}&=&\frac{k}{4\pi}\int  2 \alpha_1  R^a e_a +\alpha_2 \left( 2R^a k_a+T^a e_a +2i\bar{\psi}_aD\psi^a\right)+ \alpha_{0}L(\omega)  \,. \label{CShm}
\end{eqnarray}

The previous CS action is invariant under the $\widehat{\mathfrak{hm}}$ algebra, extending the Maxwell gravity theory with a massless spin-$\frac{5}{2}$ gauge field. Note that the terms along the arbitrary constants $\alpha_0$ and $\alpha_1$ are not affected by the extension and coincide with the terms appearing in the bosonic action \eqref{MaxCS}. The spin-$\frac{5}{2}$ field only appears in the term proportional to the $\alpha_2$ constant. Then, the action can be seen as an "exotic" hypersymmetric gravity theory which does not require the presence of spin-4 gauge fields, which turns out to be as the natural Maxwellian extension of hypergravity of Aragone and Deser in \cite{Aragone:1983sz} but formulated as a genuine gauge theory. As we shall see later, there exist other hypersymmetric extensions of the Maxwell gravity but they will require the presence of spin-4 gauge fields.

For completeness, we provide with the equations of motion which are given by 
\begin{eqnarray}
 \delta \omega_a&:& \qquad \alpha_0 R^a + \alpha_1 T^a + \alpha_2  \tilde{F}^a =0\,, \notag \\
 \delta k_a &:& \qquad \alpha_2 R^a =0\,, \notag \\
 \delta e_a &:& \qquad \alpha_1 R^a + \alpha_1 T^a =0\,, \notag \\
 \delta \bar{\psi}_a &:& \qquad \alpha_2 D\psi^a=0\,. \label{eomhm}
\end{eqnarray}
When $\alpha_2\neq0$, the previous equations can be equivalently written as the vanishing of the curvature two-forms \eqref{curvahm}. By construction, the CS action \eqref{CShm} is invariant under the local hypersymmetry transformation laws given by
\begin{eqnarray}
\delta e^{a}&=&0\,, \qquad  \ \ \ \delta \omega^{a}=0\,, \qquad \ \ \delta k^{a}=3i\Bar{\epsilon}_{b}\Gamma^{a}\psi^{b}\,, \notag \\
\delta \psi^{a}&=&d\epsilon^{a}+\frac{3}{2}\omega^{b}\Gamma_{b}\epsilon^{a}-\omega_{b}\Gamma^{a}\epsilon^{b}\,.
\end{eqnarray}
where $\epsilon^{a}$ is the fermionic gauge parameter.

The CS action \eqref{CShm} is the simplest hypersymmetric extension of the Maxwell gravity without spin-4 gauge fields. As will be discussed, the previous hypersymmetric extension of the Maxwell algebra is not unique. However, to our knowledge, it seems that the hyper-Maxwell $\widehat{\mathfrak{hm}}$ is the only consistent way to accommodate spin-$\frac{5}{2}$ generators to the Maxwell algebra without including spin-$4$ generators. 

A different analysis can be done if we choose the vierbein $e^a$ to accompany the $Z_a$ generator (or equivalently, by performing $Z_a \leftrightarrow P_a$). The analysis will lead to a hypersymmetrization of the "Hietarinta gravity" \cite{Chernyavsky:2020fqs} which although interesting, it escapes from the scope of this work.

The extension of the hyper-Maxwell gravity theory in \eqref{CShm} can then be extended to include fermionic fields of spin-$(n+3/2)$ which become  suitably described by completely symmetric and (triple) $\Gamma$-traceless  1-form $\bar{\psi}_{a_{1}\cdots a_{n}}$ so that the action reads
\begin{align}
I_{\widehat{\mathfrak{hm}}}& =\frac{k}{4\pi}\int 2\alpha_{1}R^{a}e_{a}
 +\alpha_{2}\left(2R^{a}k_{a}+T^{a}e_{a}+2i\bar{\psi}_{a_{1}\cdots a_{n}}D\psi^{a_{1}\cdots a_{n}}\right)+\alpha_{0} L(\omega) \,,
\end{align}
with
\[
D\psi^{a_{1}\cdots a_{n}}=d\psi^{a_{1}\cdots a_{n}}+\left(n+\frac{1}{2}\right)\omega_{b}\Gamma^{b}\psi^{a_{1}\cdots a_{n}}-\omega_{b}\Gamma^{\left(a_{1}\right.}\psi^{\left.a_{2}\cdots a_{n}\right)b}\;,
\]
and being left invariant by the local hypersymmetry transformations given by
\begin{align}
\delta e^{a} & =0\,,\\
\delta\omega^{a} & =0\;,\\
\delta k^{a} & =2\left(n+\frac{1}{2}\right)i\bar{\epsilon}_{a_{1}\cdots a_{n}}\Gamma^{a}\psi^{a_{1}\cdots a_{n}}\,,\\
\delta\psi^{a_{1}\cdots a_{n}} & =D\epsilon^{a_{1}\cdots a_{n}}\;.
\end{align}
Thus, the latter action extends naturally  the hypergravity theory in  \cite{Aragone:1983sz} to include the Maxwell CS gravity dynamics that amounts to describing two interacting non-propagating gravitons and a spin-$(n + 3/2)$ gauge field.


\section{On the extension of the Maxwell algebra with spin-\texorpdfstring{$\frac{5}{2}$}{32} and spin-4 generators}
In this section, we present three alternative hypersymmetric extensions of the Maxwell algebra, which are obtained through the IW contraction procedure \cite{Inonu:1953sp}. As we shall see, such hyper-Maxwell algebras require the presence of spin-4 generators and appear by contracting diverse combinations of the $\mathfrak{osp}\left(1|4\right)$ and $\mathfrak{sp}\left(4\right)$ algebras. Interestingly, we show that the hyper-Maxwell algebra \eqref{hmalgebra} without spin-4 generators (transforming in an spin-3 irreducible representation of the Lorentz group) appears as a subalgebra of one of the alternative hyper-Maxwell algebras.

To start with, and in order to fix our notation, we will briefly review the $\mathfrak{osp}\left( 1|4\right) $ superalgebra.

\subsection{The \texorpdfstring{$\mathfrak{osp}\left( 1|4\right) $}{osp} superalgebra}
The $\mathfrak{osp}\left( 1|4\right) $ superalgebra with  $\mathfrak{sl}\left( 2|R\right)$ principal  embedded in $\mathfrak{sp}\left(4\right) $ is spanned by the set of generators $\left\{
T_{a},T_{abc},\mathcal{G}_{\alpha a}\right\} $, which satisfy the following non-vanishing
(anti-)commutation relations
\begin{eqnarray}
\left[ T_{a},T_{b}\right] &=&\epsilon _{~ab}^{m}T_{m}\,,  \notag \\
\left[ T_{a},T_{bcd}\right] &=&3\epsilon _{~a\left( b\right. }^{m}T_{\left.
cd\right) m}\,,  \notag \\
\left[ T_{a},\mathcal{G}_{\alpha b}\right] &=&\frac{1}{2}\left( \Gamma
_{a}\right) _{\text{ }\alpha }^{\beta }\mathcal{G}_{\beta b}+\epsilon _{abc}%
\mathcal{G}_{\beta }^{~c}\,,  \notag \\
\left[ T_{abc},T_{mnk}\right] &=&-6\left( \eta _{\left( ab\right. }\epsilon
_{~\left. c\right) \left( m\right. }^{l}\eta _{\left. nk\right) }+5\epsilon
_{~\left( m\right. |\left( a\right. }^{l}\delta _{~b}^{d}\eta _{\left.
c\right) |n}\eta _{\left. k\right) d}\right) T_{l} \notag  \\
&&+2\left( 5\epsilon _{~\left( m\right. |\left( a\right. }^{l}\delta
_{~b}^{d}T_{\left. c\right) l|n}\eta _{\left. k\right) d}-\epsilon _{~\left(
m\right. |\left( a\right. }^{l}\eta _{\left. bc\right) }T_{|\left. nk\right)
l}-\epsilon _{~\left( m\right. \left( a\right. |}^{l}T_{\left. bc\right)
l}\eta _{|\left. nk\right) }\right) \,,  \notag \\
\left[ T_{abc},\mathcal{G}_{\alpha d}\right] &=&\left( \delta _{~d}^{k}\eta
_{\left( ab\right. |}-5\eta _{d\left( a\right. }\delta _{~b|}^{k}\right)
\left( \Gamma _{|\left. c\right) }\right) _{~\alpha }^{\beta }\mathcal{G}%
_{\beta k}+\eta _{\left( ab\right. |}\left( \Gamma _{d}\right) _{~\alpha
}^{\beta }\mathcal{G}_{\beta |\left. c\right) }\,,  \notag \\
\left\{ \mathcal{G}_{\alpha a},\mathcal{G}_{\beta b}\right\} &=&\left(
T_{abc}-\frac{4}{3}\eta _{ab}T_{c}\right) \left( C\Gamma ^{c}\right)
_{\alpha \beta }+\frac{5}{3}\epsilon _{abc}C_{\alpha \beta }T^{c}+\frac{2}{3}%
T_{\left( a\right. |}\left( C\Gamma _{|\left. b\right) }\right) _{\alpha
\beta }\,.  \label{osp14}
\end{eqnarray}%
where $a,b,\dots =0,1,2$ are Lorentz indices lowered and raised with the
off-diagonal Minkowski metric $\eta _{ab}$ and $\epsilon _{\text{ }bc}^{m}$
is the three-dimensional Levi-Civita tensor. Here $T_{a}$ span the $%
\mathfrak{sl}\left( 2,R\right) $ subalgebra  and stand for the spin-2
generators, while $T_{abc}$ and the fermionic $\mathcal{G}_{\alpha a}$ generators yield respectively to spin-4 and spin$-\frac{5}{2}$ fields in the Chern-Simons theory. Let us note that $T_{abc}$ are
traceless and totally symmetric generators satisfying $\eta ^{ab}T_{abc}=0$
while $\mathcal{G}_{\alpha a}$ are $\Gamma $-traceless vector-spinor
generators satisfying $\left( \Gamma ^{a}\right) _{\text{ }\alpha }^{\beta }%
\mathcal{G}_{\beta a}=\Gamma ^{a}\mathcal{G}_{a}=0$.  In particular, the subalgebra spanned by the set $\{T_a,T_{abc}\}$ defines a $\mathfrak{sp}(4)$ algebra.

The $\mathfrak{osp}\left( 1|4\right) $ superalgebra has the following non-vanishing components of
the invariant tensor
\begin{eqnarray}
\left\langle T_{a}T_{b}\right\rangle &=&\frac{1}{2}\eta _{ab}\,,  \notag \\
\left\langle T_{abc}T_{mnk}\right\rangle &=&5\eta _{m\left( a\right.
}\eta _{b|n}\eta _{|\left. c\right) k}-3\eta _{\left( ab\right. }\eta
_{|\left. c\right) \left( m\right. }\eta _{\left. nk\right) } \,, \\
\left\langle \mathcal{G}_{\alpha a}\mathcal{G}_{\beta b}\right\rangle &=&%
\frac{2}{3}C_{\alpha \beta }\eta _{ab}-\frac{1}{3}\epsilon _{abc}\left(
C\Gamma ^{c}\right) _{\alpha \beta }\,.  \notag
\end{eqnarray}

Let us note that the hyper-Poincaré algebra with and without spin-4 generators can be obtained by considering different IW contractions of the $\mathfrak{osp}\left(1|4\right)\otimes\mathfrak{sp}\left(4\right)$ superalgebra \cite{Fuentealba:2015wza}. Here, we shall see that three alternative hyper-Maxwell algebras appear by considering the IW contractions of the $\mathfrak{osp}\left(1|4\right)\otimes\mathfrak{osp}\left(1|4\right)\otimes\mathfrak{sp}\left(4\right)$ and $\mathfrak{osp}\left(1|4\right)\otimes\mathfrak{sp}\left(4\right)\otimes\mathfrak{sp}\left(4\right)$ superalgebras.
\subsection{Hyper-Maxwell algebra with spin-4 generators}

Let us first consider the $\mathfrak{osp}\left(1|4\right)\otimes\mathfrak{osp}\left(1|4\right)\otimes\mathfrak{sp}\left(4\right)$ superalgebra. Then, one can show that the IW contraction of such structure allows to obtain a hyper-Maxwell algebra with spin-4 generators. To this end, let us consider the following redefinition of the generators
\begin{eqnarray}
J_{a}&=&T_a+T_{a}^{+}+T_{a}^{-}\,, \qquad \qquad P_{a}=\frac{1}{\ell}\left(T_a^+-T_a^-\right)\,, \qquad \quad \, Z_{a}=\frac{1}{\ell^{2}}\left(T_a^+ + T_a^- \right)\,, \notag \\
J_{abc}&=&T_{abc}+T_{abc}^{+}+T_{abc}^{-}\,, \qquad P_{abc}=\frac{1}{\ell}\left(T_{abc}^+-T_{abc}^-\right)\,, \qquad Z_{abc}=\frac{1}{\ell^{2}}\left(T_{abc}^+ + T_{abc}^- \right)\,, \notag \\
Q_{\alpha a}&=&\frac{1}{\sqrt{\ell}}\left(\mathcal{G}_{\alpha a}^{+}-i\mathcal{G}_{\alpha a}^{-} \right) \qquad \ \ \ \Sigma_{\alpha a}=\frac{1}{\sqrt{\ell^{3}}}\left(\mathcal{G}_{\alpha a}^{+}+i\mathcal{G}_{\alpha a}^{-} \right) \,, \label{redef1}
\end{eqnarray}
where $\lbrace T_a^+,T_{abc}^+,\mathcal{G}_{\alpha a}^+ \rbrace$ and $\lbrace T_a^-,T_{abc}^-,\mathcal{G}_{\alpha a}^- \rbrace$ span each one a $\mathfrak{osp}\left(1|4\right)$ superalgebra, while $\lbrace T_a, T_{abc} \rbrace$ are $\mathfrak{sp}\left(4\right)$ generators. On the other hand, $\ell$ is a length parameter related to the cosmological constant through $\Lambda \propto \pm \frac{1}{\ell^2}$. It is simple to verify that the set of generators $\lbrace J_a,P_a,Z_a,J_{abc},P_{abc},Z_{abc},Q_{\alpha a},\Sigma_{\alpha a} \rbrace$ satisfies a hypersymmetric version of the so-called AdS-Lorentz algebra \cite{Soroka:2004fj,Soroka:2006aj,Diaz:2012zza,Concha:2017nca,Concha:2019lhn}. Then, it is straightforward to show that the algebra obtained after the vanishing cosmological constant limit $\ell\rightarrow\infty$ leads to the following hypersymmetric algebra,
\begin{eqnarray}
\left[ J_{a},J_{b}\right]  &=&\epsilon _{~ab}^{m}J_{m}\,,\qquad \ \ \ \ \ \left[
J_{a},P_{b}\right] =\epsilon _{~ab}^{m}P_{m}\,,  \notag \\
\left[ J_{a},Z_{b}\right]  &=&\epsilon _{~ab}^{m}Z_{m}\,,\qquad \ \ \ \ \ \left[
P_{a},P_{b}\right] =\epsilon _{~ab}^{m}Z_{m}\,,  \notag \\
\left[ J_{a},J_{bcd}\right]  &=&3\epsilon _{~a\left( b\right. }^{m}J_{\left.
cd\right) m}\,,\quad \left[ J_{a},P_{bcd}\right] =3\epsilon _{~a\left(
b\right. }^{m}P_{\left. cd\right) m}\,,  \notag \\
\left[ P_{a},J_{bcd}\right]  &=&3\epsilon _{~a\left( b\right. }^{m}P_{\left.
cd\right) m}\,,\quad \left[ J_{a},Z_{bcd}\right] =3\epsilon _{~a\left(
b\right. }^{m}Z_{\left. cd\right) m}\,,  \notag \\
\left[ Z_{a},J_{bcd}\right]  &=&3\epsilon _{~a\left( b\right. }^{m}Z_{\left.
cd\right) m}\,,\quad \left[ P_{a},P_{bcd}\right] =3\epsilon _{~a\left(
b\right. }^{m}Z_{\left. cd\right) m}\,, \notag \\
\left[ J_{a},Q_{\alpha b}\right]  &=&\frac{1}{2}\left( \Gamma _{a}\right) _{%
\text{ }\alpha }^{\beta }Q_{\beta b}+\epsilon _{abc}Q_{\beta }^{~c}\,,
\notag \\
\left[ J_{a},\Sigma _{\alpha b}\right]  &=&\frac{1}{2}\left( \Gamma
_{a}\right) _{\text{ }\alpha }^{\beta }\Sigma _{\beta b}+\epsilon
_{abc}\Sigma _{\beta }^{~c}\,,  \notag \\
\left[ P_{a},Q_{\alpha b}\right]  &=&\frac{1}{2}\left( \Gamma _{a}\right) _{%
\text{ }\alpha }^{\beta }\Sigma _{\beta b}+\epsilon _{abc}\Sigma _{\beta
}^{~c}\,,  \label{hm4aa}
\end{eqnarray}%
\begin{eqnarray}
\left[ J_{abc},J_{mnk}\right]  &=&-6\left( \eta _{\left( ab\right. }\epsilon
_{~\left. c\right) \left( m\right. }^{l}\eta _{\left. nk\right) }+5\epsilon
_{~\left( m\right. |\left( a\right. }^{l}\delta _{~b}^{d}\eta _{\left.
c\right) |n}\eta _{\left. k\right) d}\right) J_{l}  \notag \\
&&+2\left( 5\epsilon _{~\left( m\right. |\left( a\right. }^{l}\delta
_{~b}^{d}J_{\left. c\right) l|n}\eta _{\left. k\right) d}-\epsilon _{~\left(
m\right. |\left( a\right. }^{l}\eta _{\left. bc\right) }J_{|\left. nk\right)
l}-\epsilon _{~\left( m\right. \left( a\right. |}^{l}J_{\left. bc\right)
l}\eta _{|\left. nk\right) }\right) \,,  \notag \\
\left[ J_{abc},P_{mnk}\right]  &=&-6\left( \eta _{\left( ab\right. }\epsilon
_{~\left. c\right) \left( m\right. }^{l}\eta _{\left. nk\right) }+5\epsilon
_{~\left( m\right. |\left( a\right. }^{l}\delta _{~b}^{d}\eta _{\left.
c\right) |n}\eta _{\left. k\right) d}\right) P_{l}  \notag \\
&&+2\left( 5\epsilon _{~\left( m\right. |\left( a\right. }^{l}\delta
_{~b}^{d}P_{\left. c\right) l|n}\eta _{\left. k\right) d}-\epsilon _{~\left(
m\right. |\left( a\right. }^{l}\eta _{\left. bc\right) }P_{|\left. nk\right)
l}-\epsilon _{~\left( m\right. \left( a\right. |}^{l}P_{\left. bc\right)
l}\eta _{|\left. nk\right) }\right) \,,  \notag \\
\left[ J_{abc},Z_{mnk}\right]  &=&-6\left( \eta _{\left( ab\right. }\epsilon
_{~\left. c\right) \left( m\right. }^{l}\eta _{\left. nk\right) }+5\epsilon
_{~\left( m\right. |\left( a\right. }^{l}\delta _{~b}^{d}\eta _{\left.
c\right) |n}\eta _{\left. k\right) d}\right) Z_{l} \notag \\
&&+2\left( 5\epsilon _{~\left( m\right. |\left( a\right. }^{l}\delta
_{~b}^{d}Z_{\left. c\right) l|n}\eta _{\left. k\right) d}-\epsilon _{~\left(
m\right. |\left( a\right. }^{l}\eta _{\left. bc\right) }Z_{|\left. nk\right)
l}-\epsilon _{~\left( m\right. \left( a\right. |}^{l}Z_{\left. bc\right)
l}\eta _{|\left. nk\right) }\right) \,,  \notag \\
\left[ P_{abc},P_{mnk}\right]  &=&-6\left( \eta _{\left( ab\right. }\epsilon
_{~\left. c\right) \left( m\right. }^{l}\eta _{\left. nk\right) }+5\epsilon
_{~\left( m\right. |\left( a\right. }^{l}\delta _{~b}^{d}\eta _{\left.
c\right) |n}\eta _{\left. k\right) d}\right) Z_{l}  \notag \\
&&+2\left( 5\epsilon _{~\left( m\right. |\left( a\right. }^{l}\delta
_{~b}^{d}Z_{\left. c\right) l|n}\eta _{\left. k\right) d}-\epsilon _{~\left(
m\right. |\left( a\right. }^{l}\eta _{\left. bc\right) }Z_{|\left. nk\right)
l}-\epsilon _{~\left( m\right. \left( a\right. |}^{l}Z_{\left. bc\right)
l}\eta _{|\left. nk\right) }\right) \,,  \label{hm4bb}
\end{eqnarray}%
\begin{eqnarray}
\left[ J_{abc},Q_{\alpha d}\right]  &=&\left( \delta _{~d}^{k}\eta _{\left(
ab\right. |}-5\eta _{d\left( a\right. }\delta _{~b|}^{k}\right) \left(
\Gamma _{|\left. c\right) }\right) _{~\alpha }^{\beta }Q_{\beta k}+\eta
_{\left( ab\right. |}\left( \Gamma _{d}\right) _{~\alpha }^{\beta }Q_{\beta
|\left. c\right) }\,,  \notag \\
\left[ J_{abc},\Sigma _{\alpha d}\right]  &=&\left( \delta _{~d}^{k}\eta
_{\left( ab\right. |}-5\eta _{d\left( a\right. }\delta _{~b|}^{k}\right)
\left( \Gamma _{|\left. c\right) }\right) _{~\alpha }^{\beta }\Sigma _{\beta
k}+\eta _{\left( ab\right. |}\left( \Gamma _{d}\right) _{~\alpha }^{\beta
}\Sigma _{\beta |\left. c\right) }\,,  \notag \\
\left[ P_{abc},Q_{\alpha d}\right]  &=&\left( \delta _{~d}^{k}\eta _{\left(
ab\right. |}-5\eta _{d\left( a\right. }\delta _{~b|}^{k}\right) \left(
\Gamma _{|\left. c\right) }\right) _{~\alpha }^{\beta }\Sigma _{\beta
k}+\eta _{\left( ab\right. |}\left( \Gamma _{d}\right) _{~\alpha }^{\beta
}\Sigma _{\beta |\left. c\right) }\,, \notag \\
\left\{ Q_{\alpha a},Q_{\beta b}\right\}  &=&\left( P_{abc}-\frac{4}{3}\eta
_{ab}P_{c}\right) \left( C\Gamma ^{c}\right) _{\alpha \beta }+\frac{5}{3}%
\epsilon _{abc}C_{\alpha \beta }P^{c}+\frac{2}{3}P_{\left( a\right. |}\left(
C\Gamma _{|\left. b\right) }\right) _{\alpha \beta }\,,  \notag \\
\left\{ Q_{\alpha a},\Sigma _{\beta b}\right\}  &=&\left( Z_{abc}-\frac{4}{3}%
\eta _{ab}Z_{c}\right) \left( C\Gamma ^{c}\right) _{\alpha \beta }+\frac{5}{3%
}\epsilon _{abc}C_{\alpha \beta }Z^{c}+\frac{2}{3}Z_{\left( a\right.
|}\left( C\Gamma _{|\left. b\right) }\right) _{\alpha \beta }\,.  \label{hm4cc}
\end{eqnarray}%
The new obtained algebra, that we shall denote as $\mathfrak{hm}_{\left(4\right)}$, corresponds to a hyper-Maxwell algebra in presence of the spin-4 generators $\left\{
J_{abc},P_{abc},Z_{abc}\right\} $. This hypersymmetric extension of the Maxwell algebra is characterized by two fermionic generators $Q_{\alpha a}$ and $\Sigma_{\alpha a}$. The presence of a second spinorial charge is not arbitrary but it is required to satisfy the Jacobi Identities. Let us note that the presence of a second fermionic charge has already been discussed in  the context of $D=11$ supergravity \cite{DAuria:1982uck} and superstring theory \cite{Green:1989nn}. Subsequently, it was also studied in the context of the supersymmetric extension of the Maxwell algebra in \cite{Bonanos:2009wy,Bonanos:2010fw,deAzcarraga:2014jpa,Concha:2014xfa,Concha:2014tca,Penafiel:2017wfr,Ravera:2018vra,Concha:2018ywv}. As we will see in the next section, this novel algebra will allow to construct a different Maxwell hypergravity theory in three spacetime dimensions.

\subsection{Non-standard hyper-Maxwell algebra with spin-4 generators}
A different hyper-Maxwell algebra with one vector-spinor generator can be recovered from the $\mathfrak{osp}\left(1|4\right)\otimes\mathfrak{sp}\left(4\right)\otimes\mathfrak{sp}\left(4\right)$ superalgebra. Let us first consider the following redefinition of the generators,
\begin{eqnarray}
J_{a}&=&T_a+T_{a}^{+}+T_{a}^{-}\,, \qquad \qquad P_{a}=\frac{1}{\ell}\left(T_a^+-T_a^-\right)\,, \qquad \quad \, Z_{a}=\frac{1}{\ell^{2}}\left(T_a^+ + T_a^- \right)\,, \notag \\
J_{abc}&=&T_{abc}+T_{abc}^{+}+T_{abc}^{-}\,, \ \ \ \ \ \ P_{abc}=\frac{1}{\ell}\left(T_{abc}^+-T_{abc}^-\right)\,, \ \ \ \ \ \, Z_{abc}=\frac{1}{\ell^2}\left(T_{abc}^+ + T_{abc}^- \right)\,, \notag \\
Q_{\alpha a}&=&\frac{\sqrt{2}}{\ell}\mathcal{G}_{\alpha a}  \label{redef2}
\end{eqnarray}
where the subset $\lbrace T_a^+,T_{abc}^+,\mathcal{G}_{\alpha a} \rbrace$ satisfies an $\mathfrak{osp}\left(1|4\right)$ superalgebra, while $\lbrace T_a^-, T_{abc}^- \rbrace$ and $\lbrace T_a, T_{abc} \rbrace$ span each one a $\mathfrak{sp}\left(4\right)$ algebra. After considering the above redefinition, the set of generators $\lbrace J_a,P_a,Z_a,J_{abc},P_{abc},Z_{abc},Q_{\alpha a} \rbrace$ satisfies an alternative hypersymmetric version of the AdS-Lorentz algebra \cite{Soroka:2004fj,Soroka:2006aj,Diaz:2012zza,Concha:2017nca,Concha:2019lhn}, whose flat limit leads to the following algebra
\begin{eqnarray}
\left[ J_{a},J_{b}\right]  &=&\epsilon _{~ab}^{m}J_{m}\,,\qquad \ \ \ \ \ \left[
J_{a},P_{b}\right] =\epsilon _{~ab}^{m}P_{m}\,,  \notag \\
\left[ J_{a},Z_{b}\right]  &=&\epsilon _{~ab}^{m}Z_{m}\,,\qquad \ \ \ \ \left[
P_{a},P_{b}\right] =\epsilon _{~ab}^{m}Z_{m}\,, \notag \\
\left[ J_{a},J_{bcd}\right]  &=&3\epsilon _{~a\left( b\right. }^{m}J_{\left.
cd\right) m}\,,\quad \left[ J_{a},P_{bcd}\right] =3\epsilon _{~a\left(
b\right. }^{m}P_{\left. cd\right) m}\,, \notag \\
\left[ P_{a},J_{bcd}\right] &=&3\epsilon _{~a\left( b\right. }^{m}P_{\left.
cd\right) m}\,,\quad \left[ J_{a},Z_{bcd}\right] =3\epsilon _{~a\left(
b\right. }^{m}Z_{\left. cd\right) m}\,, \notag \\
\left[ Z_{a},J_{bcd}\right]  &=&3\epsilon _{~a\left( b\right. }^{m}Z_{\left.
cd\right) m}\,,\quad \left[ P_{a},P_{bcd}\right] =3\epsilon _{~a\left(
b\right. }^{m}Z_{\left. cd\right) m}\,, \notag \\
\left[ J_{a},Q_{\alpha b}\right]  &=&\frac{1}{2}\left( \Gamma _{a}\right) _{%
\text{ }\alpha }^{\beta }Q_{\beta b}+\epsilon _{abc}Q_{\beta }^{~c}\,, \label{hmaa}
\end{eqnarray}
\begin{eqnarray}
\left[ J_{abc},J_{mnk}\right]  &=&-6\left( \eta _{\left( ab\right. }\epsilon
_{~\left. c\right) \left( m\right. }^{l}\eta _{\left. nk\right) }+5\epsilon
_{~\left( m\right. |\left( a\right. }^{l}\delta _{~b}^{d}\eta _{\left.
c\right) |n}\eta _{\left. k\right) d}\right) J_{l}  \notag \\
&&+2\left( 5\epsilon _{~\left( m\right. |\left( a\right. }^{l}\delta
_{~b}^{d}J_{\left. c\right) l|n}\eta _{\left. k\right) d}-\epsilon _{~\left(
m\right. |\left( a\right. }^{l}\eta _{\left. bc\right) }J_{|\left. nk\right)
l}-\epsilon _{~\left( m\right. \left( a\right. |}^{l}J_{\left. bc\right)
l}\eta _{|\left. nk\right) }\right) \,,  \notag \\
\left[ J_{abc},P_{mnk}\right]  &=&-6\left( \eta _{\left( ab\right. }\epsilon
_{~\left. c\right) \left( m\right. }^{l}\eta _{\left. nk\right) }+5\epsilon
_{~\left( m\right. |\left( a\right. }^{l}\delta _{~b}^{d}\eta _{\left.
c\right) |n}\eta _{\left. k\right) d}\right) P_{l}  \notag \\
&&+2\left( 5\epsilon _{~\left( m\right. |\left( a\right. }^{l}\delta
_{~b}^{d}P_{\left. c\right) l|n}\eta _{\left. k\right) d}-\epsilon _{~\left(
m\right. |\left( a\right. }^{l}\eta _{\left. bc\right) }P_{|\left. nk\right)
l}-\epsilon _{~\left( m\right. \left( a\right. |}^{l}P_{\left. bc\right)
l}\eta _{|\left. nk\right) }\right) \,,  \notag \\
\left[ J_{abc},Z_{mnk}\right]  &=&-6\left( \eta _{\left( ab\right. }\epsilon
_{~\left. c\right) \left( m\right. }^{l}\eta _{\left. nk\right) }+5\epsilon
_{~\left( m\right. |\left( a\right. }^{l}\delta _{~b}^{d}\eta _{\left.
c\right) |n}\eta _{\left. k\right) d}\right) Z_{l} \notag \\
&&+2\left( 5\epsilon _{~\left( m\right. |\left( a\right. }^{l}\delta
_{~b}^{d}Z_{\left. c\right) l|n}\eta _{\left. k\right) d}-\epsilon _{~\left(
m\right. |\left( a\right. }^{l}\eta _{\left. bc\right) }Z_{|\left. nk\right)
l}-\epsilon _{~\left( m\right. \left( a\right. |}^{l}Z_{\left. bc\right)
l}\eta _{|\left. nk\right) }\right) \,,  \notag \\
\left[ P_{abc},P_{mnk}\right]  &=&-6\left( \eta _{\left( ab\right. }\epsilon
_{~\left. c\right) \left( m\right. }^{l}\eta _{\left. nk\right) }+5\epsilon
_{~\left( m\right. |\left( a\right. }^{l}\delta _{~b}^{d}\eta _{\left.
c\right) |n}\eta _{\left. k\right) d}\right) Z_{l}  \notag \\
&&+2\left( 5\epsilon _{~\left( m\right. |\left( a\right. }^{l}\delta
_{~b}^{d}Z_{\left. c\right) l|n}\eta _{\left. k\right) d}-\epsilon _{~\left(
m\right. |\left( a\right. }^{l}\eta _{\left. bc\right) }Z_{|\left. nk\right)
l}-\epsilon _{~\left( m\right. \left( a\right. |}^{l}Z_{\left. bc\right)
l}\eta _{|\left. nk\right) }\right)\,,\label{hmbb}
\end{eqnarray}%
\begin{eqnarray}
\left[ J_{abc},Q_{\alpha d}\right]  &=&\left( \delta _{~d}^{k}\eta _{\left(
ab\right. |}-5\eta _{d\left( a\right. }\delta _{~b|}^{k}\right) \left(
\Gamma _{|\left. c\right) }\right) _{~\alpha }^{\beta }Q_{\beta k}+\eta
_{\left( ab\right. |}\left( \Gamma _{d}\right) _{~\alpha }^{\beta }Q_{\beta
|\left. c\right) }\,,  \notag \\
\left\{ Q_{\alpha a},Q_{\beta b}\right\}  &=&\left(Z_{abc}-\frac{4}{3}\eta
_{ab}Z_{c}\right) \left( C\Gamma ^{c}\right) _{\alpha \beta }+\frac{5}{3}%
\epsilon _{abc}C_{\alpha \beta }Z^{c}+\frac{2}{3}Z_{\left( a\right.|}\left(
C\Gamma _{|\left. b\right) }\right) _{\alpha \beta }\,. 
 \label{hmcc}
\end{eqnarray}%
This algebra, which we will denote as $\widetilde{\mathfrak{hm}}_{(4)}$, corresponds to a non-standard hyper-Maxwell algebra where $P_{abc}$ and $J_{abc}$ transform in a spin-3
irreducible  representation  of  the  Lorentz  group. We refer to this algebra as non-standard since the translational generators $P_{a}$ are not expressed as bilinear expressions of fermionic generators $Q_{\alpha a}$. As we will see in the next section, this feature will imply that the hypersymmetric CS action based on this algebra, shall describe an exotic hypersymmetric action. This is analogous to what happens in the case of the non-standard Maxwell superalgebra introduced in \cite{Lukierski:2010dy} and to the simplest hyper-Maxwell algebra obtained in the previous section. 

An alternative non-standard hyper-Maxwell algebra can also be obtained from the $\mathfrak{osp}\left(1|4\right)\otimes\mathfrak{sp}\left(4\right)\otimes\mathfrak{sp}\left(4\right)$ superalgebra by considering a different redefinition of the $\mathfrak{osp}\left(1|4\right)$ and $\mathfrak{sp}\left(4\right)$ generators:
\begin{eqnarray}
J_{a}&=&T_a+T_{a}^{+}+T_{a}^{-}\,, \qquad \qquad \ \ \ \ P_{a}=\frac{1}{\ell}\left(T_a^+-T_a^-\right)\,, \qquad \quad \, Z_{a}=\frac{1}{\ell^{2}}\left(T_a^+ + T_a^- \right)\,, \notag \\
J_{abc}&=&\frac{1}{\ell}\left(T_{abc}+T_{abc}^{+}+T_{abc}^{-}\right)\,, \ \ \ \ P_{abc}=\frac{1}{\ell}\left(T_{abc}^+-T_{abc}^-\right)\,, \ \ \ \ \ \, Z_{abc}=\frac{1}{\ell}\left(T_{abc}^+ + T_{abc}^- \right)\,, \notag \\
Q_{\alpha a}&=&\frac{\sqrt{2}}{\ell}\mathcal{G}_{\alpha a}.  \label{redef3}
\end{eqnarray}
Such redefinition differs from the previous one \eqref{redef2} at the level of the length parameter $\ell$. The set of generators $\lbrace J_a,P_a,Z_a,J_{abc},P_{abc},Z_{abc},Q_{\alpha a} \rbrace$ satisfies a different hypersymmetric version of the AdS-Lorentz algebra. In this case, the flat limit $\ell\rightarrow\infty$ reproduces a different non-standard hyper-Maxwell algebra, whose non-vanishing (anti)-commutators read
\begin{eqnarray}
\left[ J_{a},J_{b}\right] &=&\epsilon _{~ab}^{m}J_{m}\,,\qquad \ \ \ \ \left[
J_{a},P_{b}\right] =\epsilon _{~ab}^{m}P_{m}\,, \notag \\
\left[ J_{a},Z_{b}\right] &=&\epsilon _{~ab}^{m}Z_{m}\,,\qquad \ \ \ \ \left[
P_{a},P_{b}\right] =\epsilon _{~ab}^{m}Z_{m}\,, \notag \\
\left[ J_{a},J_{bcd}\right]  &=&3\epsilon _{~a\left( b\right. }^{m}J_{\left.
cd\right) m}\,,\quad \left[ J_{a},P_{bcd}\right] =3\epsilon _{~a\left(
b\right. }^{m}P_{\left. cd\right) m}\,, \notag \\
 \left[ J_{a},Z_{bcd}\right]&=&3\epsilon _{~a\left(
b\right. }^{m}Z_{\left. cd\right) m}\,, \notag \\
\left[ J_{a},Q_{\alpha b}\right] &=&\frac{1}{2}\left( \Gamma _{a}\right) _{
\text{ }\alpha }^{\beta }Q_{\beta b}+\epsilon _{abc}Q_{\beta }^{~c}\,, \notag \\
\left[ J_{abc},Z_{mnk}\right]  &=&-6\left( \eta _{\left( ab\right. }\epsilon
_{~\left. c\right) \left( m\right. }^{l}\eta _{\left. nk\right) }+5\epsilon
_{~\left( m\right. |\left( a\right. }^{l}\delta _{~b}^{d}\eta _{\left.
c\right) |n}\eta _{\left. k\right) d}\right) Z_{l}\,,  \notag \\
\left[ P_{abc},P_{mnk}\right]  &=&-6\left( \eta _{\left( ab\right. }\epsilon
_{~\left. c\right) \left( m\right. }^{l}\eta _{\left. nk\right) }+5\epsilon
_{~\left( m\right. |\left( a\right. }^{l}\delta _{~b}^{d}\eta _{\left.
c\right) |n}\eta _{\left. k\right) d}\right) Z_{l}\,, \notag \\
\left[ Z_{abc},Z_{mnk}\right]  &=&-6\left( \eta _{\left( ab\right. }\epsilon
_{~\left. c\right) \left( m\right. }^{l}\eta _{\left. nk\right) }+5\epsilon
_{~\left( m\right. |\left( a\right. }^{l}\delta _{~b}^{d}\eta _{\left.
c\right) |n}\eta _{\left. k\right) d}\right) Z_{l}\,, \notag \\
\left\{ Q_{\alpha a},Q_{\beta b}\right\} &=& -\frac{4}{3}\eta
_{ab}Z_{c} \left( C\Gamma ^{c}\right) _{\alpha \beta }+\frac{5}{3}
\epsilon _{abc}C_{\alpha \beta }Z^{c}+\frac{2}{3}Z_{\left( a\right. |}\left(
C\Gamma _{|\left. b\right) }\right) _{\alpha \beta }\,. \label{hmd}
\end{eqnarray}

Interestingly, one can see that the subset  $\lbrace J_a, P_a, Z_a, Q_{\alpha a}\rbrace$ defines a hyper-Maxwell subalgebra without spin-4 generators which coincides with the simplest hyper-Maxwell algebra $\widehat{\mathfrak{hm}}$ obtained previously \eqref{hmalgebra}.


\section{Hypersymmetric extension of the Maxwell gravity theory with spin-4 gauge fields}\label{sec5}
In this section, we construct the corresponding three-dimensional CS theories invariant under the hyper-Maxwell algebra $\mathfrak{hm}_{\left(4\right)}$ with spin-$4$ generators. As we shall see, we shall require not only the presence of spin-$4$ gauge fields but also the inclusion of a second Majorana spin-$\frac{5}{2}$ gauge field. For completeness, we also present the construction of a non-standard hyper-Maxwell CS gravity based on the $\widetilde{\mathfrak{hm}}_{\left(4\right)}$ hyper-algebra.

\subsection{\texorpdfstring{$\mathfrak{hm}_{(4)}$}{hm4} Chern-Simons hypergravity}
Let us first consider the CS hypergravity theory invariant under the hyper-Maxwell algebra $\mathfrak{hm}_{(4)}$ \eqref{hm4aa}-\eqref{hm4cc}, being spanned by the set $\{P_{a}, J_{a}, Z_{a},P_{abc}, J_{abc}, Z_{abc},Q^{a}_{\alpha},\Sigma^{a}_{\alpha}\}$.  
The CS action can be constructed from the gauge field 
\begin{equation}
    A=e^{a}P_{a}+\omega^{a}J_{a}+k^{a}Z_{a}+e^{abc}P_{abc}+\omega^{abc}J_{abc}+k^{abc}Z_{abc}+\Bar{\psi}^{a}Q_{a}+\Bar{\xi}^{a}\Sigma_{a}\,,\label{oneformhm4}
\end{equation}
whose components are the dreibein, the (dualized) spin connection, the gravitational Maxwell field, three spin-4 gauge fields, and two Majorana spin-$\frac{5}{2}$ gauge fields.  The corresponding curvature two-form reads
\begin{equation}
    F_{\mathfrak{hm}_{(4)}}=\hat{\mathcal{T}}^{a}P_{a}+\mathcal{R}^{a}J_{a}+\mathcal{F}^{a}_{(\xi)}Z_{a}+\hat{\mathcal{T}}^{abc}P_{abc}+\mathcal{R}^{abc}J_{abc}+\mathcal{F}^{abc}_{(\xi)}Z_{abc}+\hat{D}\Bar{\psi}^{a}Q_{a}+D\Bar{\xi}^{a}\Sigma_{a}\,, \label{Fhm4}
\end{equation}
where
\begin{eqnarray}
\hat{\mathcal{T}}^{a}&=&T^a+30\epsilon^{a}_{\ bc}\omega^{bmn}e^{c}_{\ mn}-\frac{3}{2}i\Bar{\psi}_{b}\Gamma^{a}\psi^{b}\,, \notag \\ 
\mathcal{R}^{a}&=&R^a+15\epsilon^{a}_{\ bc}\omega^{bmn}\omega^{c}_{\ mn}\,, \notag \\
\mathcal{F}^{a}_{(\xi)}&=&F^a+30\epsilon^{a}_{\ bc}\omega^{bmn}k^{c}_{\ mn}+15\epsilon^{a}_{\ bc}e^{bmn}e^{c}_{\ mn}-3i\Bar{\psi}_{b}\Gamma^{a}\xi^{b}\,, \notag \\
\hat{\mathcal{T}}^{abc}&=&de^{abc}-10\epsilon^{(a}_{\ mn}\omega^{m k|b}e^{c)n}_{\ \ \ k}+3\epsilon^{(a}_{\ mn}e^{m}\omega^{n|bc)}+3\epsilon^{(a}_{\ mn}\omega^{m}e^{n|bc)}+\frac{i}{2}\Bar{\psi}^{(a}\Gamma^{|b}\psi^{c)}\,, \notag \\
\mathcal{R}^{abc}&=&d\omega^{abc}-5\epsilon^{(a}_{\ mn}\omega^{m k|b}\omega^{c)n}_{\ \ \ k}+3\epsilon^{(a}_{\ mn}\omega^{m}\omega^{n|bc)}\,, \notag \\
\mathcal{F}^{abc}_{(\xi)}&=&dk^{abc}-10\epsilon^{(a}_{\ mn}\omega^{m k|b}k^{c)n}_{\ \ \ k}-5\epsilon^{(a}_{\ mn}e^{m k|b}e^{c)n}_{\ \ \ k}+3\epsilon^{(a}_{\ mn}\omega^{m}k^{n|bc)}+3\epsilon^{(a}_{\ mn}k^{m}\omega^{n|bc)} \notag \\
&&+3\epsilon^{(a}_{\ mn}e^{m}e^{n|bc)}+i\Bar{\psi}^{(a}\Gamma^{|b}\xi^{c)}\,, \notag \\
\hat{D}\psi^{a}&=&d\psi^{a}+\frac{3}{2}\omega^{b}\Gamma_{b}\psi^{a}-\omega_{b}\Gamma^{a}\psi^{b}-5\omega^{bca}\Gamma_{b}\psi_{c}\,, \notag \\
D\xi^{a}&=&d\xi^{a}+\frac{3}{2}\omega^{b}\Gamma_{b}\xi^{a}-\omega_{b}\Gamma^{a}\xi^{b}-5\omega^{bca}\Gamma_{b}\xi_{c}+\frac{3}{2}e^{b}\Gamma_{b}\psi^{a}-e_{b}\Gamma^{a}\psi^{b}-5e^{bca}\Gamma_{b}\psi_{c}\,. \label{curvhm4}
\end{eqnarray}
The fermionic fields and generators are assumed to be $\Gamma$-traceless, i.e, $\Gamma^{a}\psi_{a}=\Gamma^{a}\xi_{a}=0$ and $Q^{a}\Gamma_{a}=\Sigma^{a}\Gamma_{a}=0$. The $\mathfrak{hm}_{(4)}$ hyper-algebra admits the following non-vanishing components of an invariant bilinear form
\begin{eqnarray}
\langle J_{a}J_{b} \rangle&=&\alpha_{0}\eta_{ab}\,,\qquad \ \ \langle J_{a}P_{b} \rangle=\alpha_{1}\eta_{ab}\,, \notag \\
 \langle P_{a}P_{b} \rangle&=&\alpha_{2}\eta_{ab}\,,\qquad \ \ \langle J_{a}Z_{b} \rangle=\alpha_{2}\eta_{ab}\,, \notag \\
 \langle J_{abc}J_{mnk} \rangle&=&2\alpha_{0}\left(5 \eta_{m(a}\eta_{b|n}\eta_{|c)k}-3\eta_{(ab|}\eta_{|c)}(m|\eta_{|nk)} \right)\,,\notag \\
 \langle J_{abc}P_{mnk} \rangle&=&2\alpha_{1}\left(5 \eta_{m(a}\eta_{b|n}\eta_{|c)k}-3\eta_{(ab|}\eta_{|c)}(m|\eta_{|nk)} \right)\,,\notag \\
 \langle J_{abc}Z_{mnk} \rangle&=&2\alpha_{2}\left(5 \eta_{m(a}\eta_{b|n}\eta_{|c)k}-3\eta_{(ab|}\eta_{|c)}(m|\eta_{|nk)} \right)\,,\notag \\
 \langle P_{abc}P_{mnk} \rangle&=&2\alpha_{2}\left(5 \eta_{m(a}\eta_{b|n}\eta_{|c)k}-3\eta_{(ab|}\eta_{|c)}(m|\eta_{|nk)} \right)\,,\notag \\
 \langle Q_{\alpha a}Q_{\beta b}\rangle&=&\frac{2\alpha_{1}}{3}\left(2C_{\alpha \beta} \eta_{ab}-\epsilon_{abc}\left(C\Gamma^{c}\right)_{\alpha\beta}\right)\,, \notag \\
 \langle Q_{\alpha a}\Sigma_{\beta b}\rangle &=&\frac{2\alpha_{2}}{3}\left(2C_{\alpha \beta} \eta_{ab}-\epsilon_{abc}\left(C\Gamma^{c}\right)_{\alpha\beta}\right)\,.\label{invhm4}
\end{eqnarray}
 Here $\alpha _{0}$, $\alpha _{1}$ and $\alpha _{2}$ are arbitrary constants which are related to the $\mathfrak{osp}\left(1|4\right)$ and $\mathfrak{sp}\left(4\right)$ constants given by $\left(\nu,\rho\right)$ and $\mu$, respectively. Indeed, the invariant tensor \eqref{invhm4} can be obtained from the invariant tensor of the  $\mathfrak{osp}\left(1|4\right)\otimes\mathfrak{osp}\left(1|4\right)\otimes\mathfrak{sp}\left(4\right)$ superalgebra by considering the redefinition \eqref{redef1} along with
  \begin{equation}
 \alpha_0=\frac{\mu+\nu+\rho}{2}\,,\quad \alpha_1=\frac{\nu-\rho}{2\ell}\,, 
 \quad \alpha_2=\frac{\nu+\rho}{2\ell^2}\,, \label{redefc}
 \end{equation}
 
 and the flat limit $\ell\rightarrow\infty$.
 
 Then, replacing the gauge connection one-form \eqref{oneformhm4} and the invariant tensor \eqref{invhm4} in the general expression for a CS action \eqref{CS}, it reduces, up to a surface term, to
   \begin{eqnarray}
     I_{\mathfrak{hm}_{(4)}}&=&\frac{k}{4\pi}\int  \alpha_{1}\left(2\mathcal{R}^{a}e_{a}+20\mathcal{R}^{abc}e_{abc}+2i\Bar{\psi}_{a}\hat{D}\psi^{a}\right) \notag \\
     &&+\alpha_{2}\left(2\mathcal{R}^{a}k_{a}+e^{a}\mathcal{T}_{a}+20\mathcal{R}^{abc}k_{abc}+10 e^{abc}\mathcal{T}_{abc}+2i\Bar{\psi}_{a}D\xi^{a}+2i\Bar{\xi}_{a}\hat{D}\psi^{a}\right) +
      \alpha_{0} L(\Omega) \,, \notag \\ \label{CShm4}
 \end{eqnarray}
where $\mathcal{T}^a$ and $\mathcal{T}^{abc}$ are given by
\begin{eqnarray}
\mathcal{T}^{a}&=&T^a+30\epsilon^{a}_{\ bc}\omega^{bmn}e^{c}_{\ mn}\,,  \nonumber \\ 
\mathcal{T}^{abc}&=&de^{abc}-10\epsilon^{(a}_{\ mn}\omega^{m k|b}e^{c)n}_{\ \ \ k}+3\epsilon^{(a}_{\ mn}e^{m}\omega^{n|bc)}+3\epsilon^{(a}_{\ mn}\omega^{m}e^{n|bc)}\,,
\label{Ts}
\end{eqnarray}
and
\begin{equation}
    L(\Omega) = L(\omega) +10\left(d\omega^{abc}-\frac{10}{3}\epsilon^{a}_{\ mn}\omega^{mkb}\omega^{cn}_{\ \ k}+3\epsilon^{a}_{\ mn}\omega^{m}\omega^{nbc}\right)\omega_{abc}\,,
\end{equation}
with $L\left(\omega\right)$ being the Lorentz-CS form \eqref{LorCS}.

The CS action \eqref{CShm4} is invariant under the $\mathfrak{hm}_{(4)}$ hyper-algebra and is split into three independent sectors proportional to the three arbitrary constants. The piece along $\alpha_{0}$ contains the parity-odd Lorentz CS term \cite{Achucarro:1987vz,Witten:1988hc} together with its spin-$4$ version. Along the $\alpha_1$ constant appears the Einstein-Hilbert term, contributions of the spin-4 fields and a fermionic term. The term along $\alpha_{2}$ extends the Maxwell gravity with fermionic terms and also has spin-4 fields contributions. It is straightforward to see that the previous CS action reduces to the Maxwell CS gravity \eqref{MaxCS} when the spin-$4$ and spin-$\frac{5}{2}$ fields are switch-off. Let us notice that, unlike hyper-Poincaré gravity \cite{Fuentealba:2015jma}, the presence of a second spinor gauge field  $\xi^{a}$ is required to ensure the proper extension of the Maxwell gravity with spin-$\frac{5}{2}$ and spin-$4$ gauge fields.

The field equations for $\alpha_2\neq0$ are given by the vanishing of the curvature 2-form in \eqref{Fhm4}, i.e., $F_{\mathfrak{hm}_{(4)}}=0$, whose components transform covariantly with respect to the hypersymmetry
transformation laws
\begin{eqnarray}
\delta e^{a}&=&3i\Bar{\epsilon}_{b}\Gamma^{a}\psi^{b}\,, \qquad  \ \ \ \delta \omega^{a}=0\,, \qquad \ \ \delta k^{a}=3i\Bar{\epsilon}_{b}\Gamma^{a}\xi^{b}+3i\Bar{\varrho}_{b}\Gamma^{a}\psi^{b}\,, \notag \\
\delta e^{abc}&=&-i\Bar{\epsilon}^{a}\Gamma^{b}\psi^{c}\,, \qquad  \delta \omega^{abc}=0\,, \qquad \delta k^{abc}=-i\Bar{\varrho}^{a}\Gamma^{b}\psi^{c}-i\Bar{\epsilon}^{a}\Gamma^{b}\xi^{c}\,, \notag \\
\delta \psi^{a}&=&d\epsilon^{a}+\frac{3}{2}\omega^{b}\Gamma_{b}\epsilon^{a}-\omega_{b}\Gamma^{a}\epsilon^{b}-5\omega^{bca}\Gamma_{b}\epsilon_{c}\,, \notag \\
\delta \xi^{a}&=& d\varrho^{a}+\frac{3}{2}\omega^{b}\Gamma_{b}\varrho^{a}-\omega_{b}\Gamma^{a}\varrho^{b}-5\omega^{bca}\Gamma_{b}\varrho_{c}+\frac{3}{2}e^{b}\Gamma_{b}\epsilon^{a}-e_{b}\Gamma^{a}\epsilon^{b}-5e^{bca}\Gamma_{b}\epsilon_{c}\,.
\end{eqnarray}
where $\epsilon^{a}$ and $\varrho^{a}$ are the fermionic gauge parameters related to the respective $Q_{a}$ and $\Sigma_{a}$ generators.

The novel hypergravity theory can be seen as a Maxwellian generalization of the hyper-Poincaré gravity in presence of spin-4 gauge fields. In particular, the hyper-Poincaré CS action coupled to spin-4 gauge fields and its exotic counterpart appear as particular sub-cases along the $\alpha_1$ and $\alpha_0$ constants, respectively. Let us note that, in absence of spin-4 gauge fields, the term along $\alpha_1$ constant corresponds to the usual hypergravity introduced in \cite{Aragone:1983sz,Fuentealba:2015jma}. Although, the $\mathfrak{hm}_{\left(4\right)}$ hyper-algebra can be seen as an enlargement and deformation of the hyper-Poincaré symmetry, the present hypergravity theory does not contain a cosmological constant term.

\subsection{Non-standard hypersymmetric Maxwell Chern-Simons action}
Now we shall construct the CS action invariant under the non-standard hyper-Maxwell algebra in presence of spin-4 generators given by \eqref{hmaa}-\eqref{hmcc}, being spanned by $\{P_{a}, J_{a}, Z_{a},J_{abc},P_{abc},Z_{abc},Q^{a}_{\alpha}\}$.  
The gauge field one-form is given by
\begin{equation}
    A=e^{a}P_{a}+\omega^{a}J_{a}+k^{a}Z_{a}+e^{abc}P_{abc}+\omega^{abc}J_{abc}+k^{abc}Z_{abc}+\Bar{\psi}^{a}Q_{a}\,,\label{onehm}
\end{equation}
whose components are the dreibein, the (dualized) spin connection, the gravitational Maxwell field, three spin-4 gauge fields, and one Majorana spin-$5/2$ field. The curvature two-form is given in this case by
\begin{equation}
    F_{\widetilde{\mathfrak{hm}}_{\left(4\right)}}=\mathcal{T}^{a}P_{a}+\mathcal{R}^{a}J_{a}+\mathcal{F}^{a}_{(\psi)}Z_{a}+\mathcal{T}^{abc}P_{abc}+\mathcal{R}^{abc}J_{abc}+\mathcal{F}^{abc}_{(\psi)}Z_{abc}+\hat{D}\Bar{\psi}^{a}Q_{a}\,,
    \label{curhmc}
\end{equation}
where $\mathcal{T}^a$, $\mathcal{T}^{abc}$ are defined in \eqref{Ts}, while $\mathcal{R}^a$, $\mathcal{R}^{abc}$, $\hat{D}\psi^a$ can be found in \eqref{curvhm4}, and
\begin{eqnarray}
\mathcal{F}^{a}_{(\psi)}&=&F^a+30\epsilon^{a}_{\ bc}\omega^{bmn}k^{c}_{\ mn}+15\epsilon^{a}_{\ bc}e^{bmn}e^{c}_{\ mn}-\frac{3}{2}i\Bar{\psi}_{b}\Gamma^{a}\psi^{b}\,, \notag \\
\mathcal{F}^{abc}_{(\psi)}&=&dk^{abc}-10\epsilon^{(a}_{\ mn}\omega^{m k|b}k^{c)n}_{\ \ \ k}-5\epsilon^{(a}_{\ mn}e^{m k|b}e^{c)n}_{\ \ \ k}+3\epsilon^{(a}_{\ mn}\omega^{m}k^{n|bc)}+3\epsilon^{(a}_{\ mn}k^{m}\omega^{n|bc)} \notag \\
&&+3\epsilon^{(a}_{\ mn}e^{m}e^{n|bc)}+\frac{i}{2} \Bar{\psi}^{(a}\Gamma^{|b}\psi^{c)}\,.  \label{curhm}
\end{eqnarray}
The algebra $\widetilde{\mathfrak{hm}}_{\left(4\right)}$ admits the following non-vanishing components of an invariant bilinear form,
\begin{eqnarray}
\langle J_{a}J_{b} \rangle&=&\alpha_{0}\eta_{ab}\,,\qquad \ \ \langle J_{a}P_{b} \rangle=\alpha_{1}\eta_{ab}\,, \notag \\
 \langle P_{a}P_{b} \rangle&=&\alpha_{2}\eta_{ab}\,,\qquad \ \ \langle J_{a}Z_{b} \rangle=\alpha_{2}\eta_{ab}\,, \notag \\
 \langle J_{abc}J_{mnk} \rangle&=&2\alpha_{0}\left(5 \eta_{m(a}\eta_{b|n}\eta_{|c)k}-3\eta_{(ab|}\eta_{|c)}(m|\eta_{|nk)} \right)\,,\notag \\
 \langle J_{abc}P_{mnk} \rangle&=&2\alpha_{1}\left(5 \eta_{m(a}\eta_{b|n}\eta_{|c)k}-3\eta_{(ab|}\eta_{|c)}(m|\eta_{|nk)} \right)\,,\notag \\
 \langle J_{abc}Z_{mnk} \rangle&=&2\alpha_{2}\left(5 \eta_{m(a}\eta_{b|n}\eta_{|c)k}-3\eta_{(ab|}\eta_{|c)}(m|\eta_{|nk)} \right)\,,\notag \\
 \langle P_{abc}P_{mnk} \rangle&=&2\alpha_{2}\left(5 \eta_{m(a}\eta_{b|n}\eta_{|c)k}-3\eta_{(ab|}\eta_{|c)}(m|\eta_{|nk)} \right)\,,\notag \\
 \langle Q_{\alpha a}Q_{\beta b}\rangle&=&2\alpha_{2}\left(\frac{2}{3}C_{\alpha \beta} \eta_{ab}-\frac{1}{3}\epsilon_{abc}\left(C\Gamma^{c}\right)_{\alpha\beta}\right)\,.\label{invhm}
\end{eqnarray}
Here $\alpha _{0}$, $\alpha _{1}$ and $\alpha _{2}$ are arbitrary constants which are related to the $\mathfrak{osp}\left(1|4\right)$ and $\mathfrak{sp}\left(4\right)$ constants given by $\nu$ and $\left(\mu,\rho\right)$, respectively. Indeed, the invariant tensor \eqref{invhm} can be obtained from the invariant tensor of the  $\mathfrak{osp}\left(1|4\right)\otimes\mathfrak{sp}\left(4\right)\otimes\mathfrak{sp}\left(4\right)$ superalgebra by considering the redefinition \eqref{redef2} along with \eqref{redefc} and the flat limit $\ell\rightarrow\infty$.

Then, the CS action is given, up to boundary terms, by
\begin{eqnarray}
     I_{\widetilde{\mathfrak{hm}}_{(4)}}&=&\frac{k}{4\pi}\int 
     \alpha_{1}\left(2\mathcal{R}^{a}e_{a}+20\mathcal{R}^{abc}e_{abc}\right) \notag \\
     &&+\alpha_{2}\left(2\mathcal{R}^{a}k_{a}+e^{a}\mathcal{T}_{a}+20\mathcal{R}^{abc}k_{abc}+10 e^{abc}\mathcal{T}_{abc}+2i\Bar{\psi}_{a}\hat{D}\psi^{a}\right) +\alpha_{0} L(\Omega) \,. \label{CShmm}
 \end{eqnarray}
The CS action based on the $\widetilde{\mathfrak{hm}}_{\left(4\right)}$ hyper-algebra describes an exotic hypersymmetric gravity  theory. One can see that the spin-$\frac{5}{2}$ gauge fields do not contribute to the Einstein-Hilbert sector. This is mainly due to the structure of the $\widetilde{\mathfrak{hm}}_{\left(4\right)}$ hyper-algebra \eqref{hmaa}-\eqref{hmcc} in which $\{Q,Q\} \not\sim P$. Let us note that the same behavior appears in the simplest hyper-Maxwell gravity \eqref{CShm} and in the non-standard supersymmetric extension of the Maxwell gravity \cite{Concha:2020atg}. It would be worth it to study the Hietarinta version \cite{Bansal:2018qyz,Chernyavsky:2019hyp,concha:2021aaa} of the non-standard hyper-Maxwell CS action \eqref{CShmm}, in which the role of the dreibein and the gravitational Maxwell gauge field are interchanged, and its physical implications.
The field equations for $\alpha_2\neq0$ are given by the vanishing of the curvature 2-form in \eqref{curhmc}, i.e., $F_{\widetilde{\mathfrak{hm}}_{\left(4\right)}}=0$. In the present case, the curvatures transform covariantly under the hypersymmetry transformation laws:
\begin{eqnarray}
\delta e^{a}&=&0\,, \qquad  \ \ \ \delta \omega^{a}=0\,, \qquad \ \ \delta k^{a}=3i\Bar{\epsilon}_{b}\Gamma^{a}\psi^{b}\,, \notag \\
\delta e^{abc}&=&0\,, \qquad  \  \delta \omega^{abc}=0\,, \qquad \delta k^{abc}=-i\Bar{\epsilon}^{a}\Gamma^{b}\psi^{c}\,, \notag \\
\delta \psi^{a}&=&d\epsilon^{a}+\frac{3}{2}\omega^{b}\Gamma_{b}\epsilon^{a}-\omega_{b}\Gamma^{a}\epsilon^{b}-5\omega^{bca}\Gamma_{b}\epsilon_{c}\,,
\end{eqnarray}
where $\epsilon^{a}$ is the fermionic gauge parameter related to $Q_{a}$.


\section{Concluding remarks}

In this work, we have presented a consistent way of coupling three-dimensional Maxwell CS gravity theory with a spin-$\frac{5}{2}$ gauge field. To this end, we have constructed the simplest hypersymmetric extension of the Maxwell algebra, which was carried out by the consistent insertion of new fermionic generators into the three dimensional Maxwell group which transform in an spin-$\frac{3}{2}$ irreducible representation of the Lorentz group. The respective CS theory invariant under the aforesaid hyper-Maxwell algebra was then introduced and can be seen as an exotic hypersymmetric gravity theory that extend the hypergravity of Aragone and Deser \cite{Aragone:1983sz} that  allowed us  to write its corresponding extension to includes massless gauge fields of spin-$(n+3/2)$. Interestingly, it was also shown that the hypersymmetric extension of the Maxwell algebra is not unique but can be further extended to include spin-$4$ generators that are accompanied by  additional fermionic spin-$\frac{3}{2}$ ones. Indeed, it was shown explicitly that three different hyper-Maxwell algebras can be derived by considering the IW contraction of diverse combinations of the $\mathfrak{osp}\left(1|4\right)$ and the $\mathfrak{sp}\left(4\right)$ algebras. The first of these superalgebras obtained here \eqref{hm4aa}-\eqref{hm4cc} allowed to construct a consistent CS hypergravity theory that amounts to endowing Maxwell gravity with interacting  massless spin-$4$ fields and two independent fermionic gauge fields of spin-$\frac{5}{2}$. A second alternative hypersymmetric algebra \eqref{hmaa}-\eqref{hmcc}, named here as a non-standard algebra, allowed to construct the non-standard hypersymmetric Maxwell CS action being characterized by having just one spin-$\frac{5}{2}$ fermionic gauge field. Notably, in the third case \eqref{hmd} the spin-$4$ gauge fields can be consistently truncated which goes along with the fact that the set of spin-$4$ generators form indeed an subalgebra, so the simplest hyper-Maxwell gravity theory that includes a single massless  spin-$\frac{5}{2}$ gauge field is effectively recovered.

Note that for our simplest hyper-Maxwell Chern-Simons gravity, similar to the case of the Poincaré hypergravity \cite{Fuentealba:2015jma}, the inclusion of additional HS generators of half-integer spin into the Maxwell algebra does not affect the  causal structure underlying Maxwell CS gravity theory, which distinguished it from the more familiar infinite-dimensional higher-spin algebras which requires to extend Lorentz  in order to include generators of higher spins $s>2$.

The results obtained here could serve as a starting point for various further studies. In particular, one could explore the possibility to include a cosmological constant to our theory. Indeed, with regard to the bosonic sector of the theory it is encouraging to known that the Maxwell gravity theory can alternatively be recovered as a vanishing cosmological constant limit of the so-called AdS-Lorentz gravity \cite{Diaz:2012zza,Concha:2018jxx,Concha:2018jjj} and the Poincaré-Lorentz gravity theory \cite{Adami:2020xkm}. Then, a cosmological constant could be incorporated in the hyper-Maxwell gravity theory by constructing a CS hypergravity theory based on a hyper-AdS-Lorentz or a hyper-Poincaré-Lorentz symmetry. Although, they probably should appear as deformations of the hyper-Maxwell algebras presented here, one could expect different underlying geometries. Indeed, unlike the AdS-Lorentz gravity theory, the inclusion of a cosmological constant through the Poincaré-Lorentz symmetry implies the presence of a non-vanishing torsion \cite{Adami:2020xkm}. It would be worth it to explore the physical implications of considering a non-vanishing torsion in a hypergravity theory.

It is worth stressing that the study of suitable asymptotic symmetries for the  extensions of hyper-Maxwell theories found here become to be crucial due to their inherent topological character pointed by the absence of local bulk degrees of freedom \cite{Bunster:2014mua}. Thus, one could expect that suitable asymptotic conditions can be performed in each case along the lines of \cite{Henneaux:2013dra,Bunster:2014mua} giving raise to adequate asymptotic symmetries being canonically realized by deformations of the hyper-$\mathfrak{bms}_3$ asymptotic algebra. One might then wonder whether interesting hypersymmetry bounds can be derive from these asymptotic algebras implying prominent properties for solutions such as soliton-like as well as those with a sensible thermodynamics, see e.g., \cite{Henneaux:2015tar,Fuentealba:2015jma}. Indeed, since the hyper-Maxwell algebra appears as the IW contraction of diverse combinations of the $\mathfrak{osp}\left(1|4\right)$ and the $\mathfrak{sp}\left(4\right)$ algebras, it seems natural to expect that the respective asymptotic algebra could alternatively be recovered as a precise combination of the $W_{\left(2,\frac{5}{2},4\right)}$ and $W_{\left(2,4\right)}$ algebras (work in progress).

Another interesting aspect that deserves to be explored is the derivation of the hyper-Maxwell algebras introduced here by considering the S-expansion method. In three spacetime dimensions, the Maxwell algebra and its supersymmetric extension can be obtained by expanding the $\mathfrak{so}\left(2,1\right)$ and $\mathfrak{osp}\left(2,1\right)$ algebras, respectively \cite{Concha:2018jxx}. One could extend the S-expansion procedure at the hypersymmetric level by studying the S-expansion of the $\mathfrak{osp}\left(1|4\right)$ superalgebra considering different semigroups. It would be interesting to recover not only known algebras (as the hyper-Poincaré and the ones presented here) but also novel algebras being the hypersymmetric extensions of known (or unknown) bosonic algebras. A particular advantage of considering the S-expansion approach is that it provides us not only with the expanded (anti-)commutation relations but also with the non-vanishing components of the invariant tensor of the expanded algebra, which are essentials in the construction of a CS action.

It is well known that three-dimensional CS actions possess higher-dimensional generalizations. Then, following \cite{Fuentealba:2019bgb}, it would be worth it to extend our construction to five and higher odd spacetime dimensions. As an ending remark, let us mention that the $\mathcal{N}$-extension of our results can be done considering an appropriate IW contraction of diverse combinations of the $\mathfrak{osp}\left(M|4\right)$ and $\mathfrak{sp}\left(4\right)$ algebras. In particular, one could expect that a $\mathcal{N}=\left(M,N\right)$ extended version of the simplest hyper-Maxwell algebra presented here can be derived from the $\mathfrak{osp}\left(M|4\right)\times\mathfrak{osp}\left(N|4\right)\times\mathfrak{sp}\left(4\right)$ superalgebra. Let us note that the $\mathfrak{osp}\left(M|4\right)\times\mathfrak{osp}\left(N|4\right)$ superalgebra corresponds to the $\mathcal{N}$-extended hyper-AdS algebra studied in \cite{Henneaux:2015tar}. 

\section*{Acknowledgment}

 The work of R.C., D.T. and P.C. is partially funded by the National Agency for Research and Development ANID (ex-CONICYT) - FONDECYT grants No. 1211077, 1181031, 1181496, 1211226 and PAI grant No. 77190078. This work was supported by the Research project Code DIREG$\_$09/2020 (R.C. and P.C.) of the Universidad Católica de la Santisima Concepción, Chile. R.C. and P.C. would like to thank to the Dirección de Investigación and Vice-rectoría de Investigación of the Universidad Católica de la Santísima Concepción, Chile, for their constant support. The work of J.M. was supported by the ERC Advanced Grant “High-Spin-Grav", by FNRS-Belgium (convention FRFC PDR T.1025.14 and convention IISN 4.4503.15).


\bibliographystyle{fullsort.bst}
 
\bibliography{Maxwell_Hypergravity}

\end{document}